\documentclass[twocolumn]{jpsj2} 

\def\runauthor{Gen'ya \textsc{Sakurai} and Yoshio \textsc{Kuramoto}}

\title{%
Wavenumber Dependence of 
Multipolar Interactions 
in the Anderson Lattice}

\author{%
Gen'ya \textsc{Sakurai}\thanks{E-mail address: genya@cmpt.phys.tohoku.ac.jp} and Yoshio \textsc{Kuramoto}}

\inst{%
Department of Physics, Tohoku University, Sendai 980-8578}

\recdate{\today}

\abst{%
Multipolar interactions are derived in the orbitally degenerate Anderson lattice with a spherical Fermi surface and one conduction electron per unit cell of the simple cubic lattice. 
As the crystalline-electric-field (CEF) ground state of $f^1$ configuration, the 
four-fold degenerate $\Gamma_8$ is mainly studied.
Intersite interactions up to a sufficiently distant pair are Fourier transformed to the wavenumber space.  
For the $\Gamma_8$ case,
quadrupolar and octupolar interactions favor the staggered order with $\mib{q} = (1/2,1/2,1/2)$,
while the dipolar interaction favors an incommensurate magnetic structure with a long modulation period.
The latter is due to a Kohn anomaly which is 
found to be
sharply peaked
for interaction channels with large angular momenta. 
Implications of results are discussed for multipole orders in CeB$_6$,  and the incommensurate magnetic structure in a quasi-cubic system CeB$_2$C$_2$.
}

\kword{incommensurate magnetic order, Kohn anomaly,
Anderson lattice, RKKY interaction
}

\begin{document}
\sloppy
\maketitle
\section{Introduction}

It is a long-standing problem in magnetism to identify the microscopic origin of complicated magnetic structures in solids.  
In some cases, a complicated structure results from coupling between dipoles and orbital degrees of freedom, or the quadrupole moment.
Accordingly ordering of multipole moments higher than dipoles has attracted much attention.
In CeB$_6$ for example, a peculiar magnetic structure is realized on the background of a quadrupole (or orbital) order \cite{Effantin}.
The resultant wave number $\mib{q} =(1/4,\pm 1/4, 1/2)$ of the magnetic moment is closely related to the quadrupole order with $\mib{q} =(1/2, 1/2, 1/2)$.  Thus 
it is tempting to interpret the magnetic pattern in terms of the 
background quadrupole patterns together with the strong spin-orbit interaction.

However, a magnetic order with $\mib{q} =(1/4,\pm 1/4, 1/2)$ 
has also been found in other RB$_6$ systems including R $=$ Gd \cite{Kuwahara-GdB6}.  
Since the trivalent Gd with $4f^7$ configuration does not have the orbital degrees of freedom, the characteristic $\mib q$ should originate from structure of conduction bands.   
A model has been proposed
to explain the stability of $\mib{q} =(1/4,\pm 1/4, 1/2)$ \cite{Kuramoto-Kubo}.
The model emphasizes the interpocket polarization 
among the three nearly spherical Fermi surfaces in the Ruderman-Kittel-Kasuya-Yosida (RKKY) interaction. 

On the other hand, related compounds
RB$_2$C$_2$  (R $=$ rare earths), which crystallize in the tetragonal LaB$_2$C$_2$-type structure({\it P}4/{\it mbm}) \cite{Ohoyama3},
often exhibit an incommensurate magnetic structure \cite{Ohoyama1}.
For example, CeB$_2$C$_2$ undergoes a magnetic ordering at $T_{\textrm{N}}$ = 7.3K, and a successive transition at $T_{\textrm{t}}$ = 6.5K \cite{Onodera1}.
Below $T_{\textrm{t}}$, a long-period magnetic structure with its propagation vector $\mib{k} = (\delta, \delta, \delta')$, where $\delta = 0.16$ and $\delta' = 0.10$, is found from neutron diffraction \cite{Ohoyama1}.
Thus propagation vectors seem to be characterized by conduction bands near the Fermi level.

RB$_2$C$_2$ can be considered as a quasi-cubic system because the distance $a/\sqrt{2}$ between R ions in the $c$-plane is almost the same as the distance $c$ along the $c$-axis\cite{Ohoyama3}.
The underlying atomic composition is closely related to RB$_6$ systems. 
Namely, if one removes two boron atoms from each B$_6$ cluster, and further replaces two boron atoms by two carbon atoms to make a B$_2$C$_2$ cluster, one obtains the RB$_2$C$_2$ structure.  
According to band-structure calculation for LaB$_2$C$_2$ \cite{Harima}, 
a connected ellipsoidal 
Fermi surface is formed around the Z point in the simple tetragonal Brillouin zone (BZ).  This Fermi surface is partly confirmed by de Haas-van Alphen (dHvA) experiment \cite{Watanuki1}.
Qualitatively speaking, the ellipsoidal shape is a result of  
growth from one of three equivalent Fermi surfaces in RB$_6$.
The other two pieces shrink and disappear.

It should be clear from the examples above that the interplay of energy-band structure and the 4$f$ shell structure deserves detailed inspection.  The purpose of this paper is to derive intersite interactions of multipoles from microscopic calculation for a simple model which still keeps the realistic 4$f$ shell structure.   
The RKKY interaction should largely be determined by hybridization in the case of CeB$_6$ and related compounds.
In a previous paper, we calculated the RKKY interaction for the nearest and the next nearest neighbors taking the Anderson lattice with orbital degeneracy. 
The conduction-band states are taken as plane waves and the hybridization as spherically symmetric \cite{GS-Kuramoto}.
In this paper we extend the calculation to many different pairs to obtain the intersite interaction in the wavenumber space.
We again take the spherical Fermi surface in the simple cubic BZ (SCBZ) with the lattice constant $a_q\equiv a/\sqrt{2}$. 
The result can be mapped onto the the quasi-cubic BZ (QCBZ)
of the bace-centred tetragonal lattice with lattice constants $a$ and $c=a_q$.
We consider mainly the $\Gamma_8$ quartet of total angular momentum $j = 5/2$ as the lowest CEF states.
This quartet includes  the $m = \pm 1/2$ doublet where $m$ is the azimuthal quantum number of the total angular momentum along the $c$ axis. 
The doublet is likely to be the CEF ground state in the tetragonal CeB$_2$C$_2$ \cite{Suzuki1}.
Hence we shall also derive corresponding results for 
the $m = \pm 1/2$ doublet, and for another doublet $\Gamma_7$.  

This paper is organized as follows.
In \S 2 we briefly review the
effective intersite Hamiltonian derived previously \cite{GS-Kuramoto}.
In \S 3 multipolar couplings are calculated numerically 
for the simple cubic lattice, and are transformed to wavenumber space.
Relation between possible ordering structures and the Fermi surface is discussed with emphasis on the Kohn anomaly.
We discuss implications of results on actual materials in \S 4.
Appendix gives 
some analytic results about the polarization functions with finite angular momentum.
\section{Intersite Interactions from Real to Wavenumber Spaces}
We start from the 
spherically symmetric form of hybridization:
\begin{align}
H_{\text{hyb}} =& \sqrt{\frac{4\pi}{N}}\sum_{i \mib{k} m \sigma}
		\left[V 
f_{m\sigma}^{\dagger}(i) 
		Y_{3m}^\ast (\hat{\mib{k}})
\mathrm{e}^{\mathrm{i}\mib{k}\cdot\mib{R}_i}
c_{\mib{k}\sigma} + \text{H.c.}\right], \label{Hhyb}
\end{align}
where $N$ is the number of lattice sites, $V$ is the strength of hybridization, 
and $Y_{lm}(\Omega)$ is the spherical harmonics with solid angle $\Omega$.
A 4$f$ electron is created by 
$f_{m\sigma}^\dagger(i)$ 
at site $\mbox{\boldmath$R$}_i$ with $z$-component $m$ of the orbital angular momentum $l=3$ and spin $\sigma$.
A conduction electron with wavenumber $\mbox{\boldmath$k$}$ and spin $\sigma$ is annihilated by $c_{\mbox{\boldmath$k$}\sigma}$. 
Each Ce$^{3+}$ ion has one 4$f$ electron, and the spin-orbit coupling leads to the total angular momentum 
$j_1 = 5/2$ in the ground state.  
By using the fourth-order perturbation theory with respect to $H_{\text{hyb}}$, we obtain the multipolar intersite interaction $H_{ij}$
between $\mib{R}_i$ and $\mib{R}_j$ \cite{GS-Kuramoto}:
\begin{align}
&H_{ij} =
\frac{|V|^4}{4\pi^{\frac{7}{2}}E_a}
\sum_{\xi_i \xi_j}
\sum_{l_1, l_2, l_3 = 0}^\infty
\sum_{P_i, P_j = 0}^{2j_1}
\sum_{j_s j_t j_u j_v}^{\frac{5}{2}, \frac{7}{2}} \nonumber \\
& \times
\frac{
\Lambda(\xi_i; j_t,j_u,P_i)\Lambda(\xi_j; j_v, j_s, P_j )
}{E_{\text{exc}}(\xi_i)E_{\text{exc}}(\xi_j)}
(k_{\textrm{F}} a_q)^4Q_{l_1 l_2}(k_{\textrm{F}} R_{ij})
\nonumber \\
& \times
(-1)^{(l_1+l_2)/2-j_s -j_t -j_u - j_v - P_i} \nonumber \\
& \times [3]^2 [l_1, l_2, P_i, P_j][j_s, j_t, j_u, j_v, l_3]^{\frac{1}{2}}
\nonumber \\
& \times
\begin{pmatrix}
		\!3\! & \!l_1\! & \!3\! \\
		\!0\! & \!0\! & \!0\!
\end{pmatrix}
\begin{pmatrix}
		\!3\! & \!l_2\! & \!3\! \\
		\!0 & \!0\! & \!0\!
\end{pmatrix}
\begin{pmatrix}
		\!l_3\! & \!l_1\! & \!l_2\! \\
		\!0\! & \!0\! & \!0\!
\end{pmatrix} \nonumber \\
& \times
\begin{Bmatrix}
		\!3\! & \!3\! & \!l_1\! \\
		\!j_t\! & \!j_s\! & \frac{1}{2}
\end{Bmatrix}
\begin{Bmatrix}
		\!3\! & \!3\! & \!l_2\! \\
		\!j_v\! & \!j_u\! & \!\frac{1}{2}\!
\end{Bmatrix}
\begin{Bmatrix}
		\!P_i\! & \!j_u\! & \!j_t\! \\
		\!P_j\! & \!j_v\! & \!j_s\! \\
		\!l_3\! & \!l_2\! & \!l_1\!
\end{Bmatrix}
\nonumber \\
& \times
\frac{
\{\mib{T}^{P_i}(i)\mib{T}^{P_j}(j)\mib{Y}_{l_3}(\hat{\mib{R}}_{ij})\}_0^0
}{\langle j_1 || \mib{T}^{P_i} || j_1 \rangle \langle j_1 || \mib{T}^{P_j} || j_1 \rangle}, \label{HijITO}
\end{align}
where $k_{\textrm{F}}$ is the Fermi wavenumber, and
$\mib{R}_{ij} = \mib{R}_i-\mib{R}_j$. 
We have introduced $E_a=\hbar^2/2m_0 a_q^2$ with
$m_0$ being the effective mass of conduction electrons. 
In eq.(\ref{HijITO})  
$Q_{l_1 l_2}(x)$ is the range function tabulated in ref.\citen{GS-Kuramoto},
$E_{\textrm{exc}}(\xi)$ is the excitation energy to a state $\xi$.
The shorthand notation
$[j_1,\cdots,j_n] = (2j_1+1)
\cdots (2j_n+1)$
is used, and round or curly brackets denote Wigner's 3{\it n-j} symbols.
We have introduced
$\mib{T}^P(i)$ as the set of rank-$P$ irreducible-tensor operators at site 
$\mib{R}_i$ with $\langle j_1||\mib{T}^P||j_1 \rangle$ its reduced matrix element, 
and $\{ \cdots \}_0^0$ denotes a tensor component of rank zero, i.e., a scalar \cite{Teitelbaum}.
The factor $\Lambda(\xi;j_t,j_u,P)$ includes information of wave functions of the ground state and an excited state $\xi$.
If we take $\xi$ as the 4$f^0$ state or a 4$f^2$ state with orbital angular momentum $L$, spin $S$ and total angular momentum $J$, $\Lambda(\xi;j_t,j_u,P)$ is given by
\begin{align}
\Lambda(f^0; j_t, j_u, P) \!&=-
\langle j_1 ||f_{j_t}^\dagger ||0 \rangle
\langle j_1 ||f_{j_u}^\dagger ||0 \rangle^\ast \nonumber \\
& \times
(-1)^{j_1 + j_u - P}
\begin{Bmatrix}
		j_1 & j_1 & P \\
		j_t & j_u & 0
\end{Bmatrix}, \label{LambdaZero}
\\
\Lambda(f^2(LS)J; j_t, j_u, P) \! &=
\langle (LS)J || f_{j_u}^\dagger ||j_1 \rangle^\ast
\langle (LS)J || f_{j_t}^\dagger ||j_1 \rangle \nonumber \\
& \times (-1)^{j_1 + j_t + J + 2P}
\begin{Bmatrix}
		j_1 & j_1 & P \\
		j_t & j_u & J
\end{Bmatrix}. \label{LambdaTwo}
\end{align}

From eq.(\ref{HijITO}), 
we obtain the intersite interaction projected by $\mathcal{P}$ to low-lying CEF states: 
\begin{align}
&\sum_{ij} \mathcal{P}
H_{ij} \mathcal{P} \nonumber \\
=& \sum_{\xi \xi'}
\frac{|V|^4}{E_a E_{\mathrm{exc}}(\xi) E_{\mathrm{exc}}(\xi')}
\sum_{i j \alpha \beta}
D^{\alpha \beta}(\xi,\xi';\mib{R}_{ij})
X^{\alpha}(i) X^{\beta}(j) \nonumber \\
=&
\sum_{\xi \xi'}\frac{|V|^4}{E_a E_{\mathrm{exc}}(\xi) E_{\mathrm{exc}}(\xi')}
\sum_{\mib{q} \alpha \beta}
\tilde{D}^{\alpha \beta}(\xi,\xi'; \mib{q})
\tilde{X}^{\alpha}(-\mib{q})\tilde{X}^{\beta}(\mib{q}),
\end{align}
where 
$X^{\alpha}(i)$ is a 
multipole operator at site $\mib{R}_i$ \cite{GS-Kuramoto}, $D^{\alpha \beta}(\xi, \xi';\mib{R}_{ij})$ is the dimensionless multipolar coupling strength.
In the case of $\Gamma_8$ state, $\alpha$ and $\beta$ run from 1 to 15.
Their Fourier transforms are defined as follows:
\begin{align}
X^{\alpha}(i) &=\frac{1}{\sqrt{N}}\sum_{\mib{q}}\tilde{X}^{\alpha}(\mib{q})
\exp[-\mathrm{i}\mib{q}\cdot\mib{R}_i], \\
D^{\alpha \beta}(\xi, \xi'; \mib{R}_{ij}) &=\frac{1}{N}\sum_{\mib{q}}
\tilde{D}^{\alpha \beta}(\xi, \xi'; \mib{q})\exp[-\mathrm{i}\mib{q}\cdot \mib{R}_{ij}].
\end{align}
With the time-reversal symmetry, multipole operators are classified into electric (even) and magnetic (odd) multipole operators.
In the intersite interaction, the operators with different time-reversal symmetry do not couple.
Therefore we obtain a block-diagonalized matrix for $\tilde{D}^{\alpha \beta}(\xi, \xi'; \mib{q})$
where submatrices
$\tilde{\mib{D}}_{\nu}(\xi, \xi'; \mib{q})$ with either even ($\nu=g$) or odd ($\nu=u$) symmetry under the time reversal can be defined.
Then we diagonalize the submatrices for each $\mib{q}$.
We note that comparison of relative strength of resultant multipolar couplings 
is physically relevant because all $X^{\alpha}(i)$'s are normalized
in the same way.
In the next section, we diagonalize $\tilde{\mib{D}}_{\nu}(\xi, \xi'; \mib{q})$ numerically and discuss 
possible ordering structures.

\section{Numerical Results of Intersite Interactions}
%
To obtain the wavenumber dependence of intersite interactions, we first calculate $\mib{D}_{\nu}(\xi, \xi'; \mib{R}_{0j})$ numerically for a sufficiently large number of sites $\mib{R}_j$.
As for the intermediate states,  both 4$f^0$ and 4$f^2$ Hund's-rule ground states are considered, and they are referred to $f^0$ and $f_{\text{H}}^2$ respectively.
We define three cases depending on the intermediate states at 
$\mib{R}_0$ and $\mib{R}_j$:
 (a) $f^0$ intermediate states on 
 both sites, 
(b) $f_{\text{H}}^2$ intermediate states on both sites, and \\
(c) $f^0$ on one site and $f_{\text{H}}^2$ on the other sites. 
We 
exclude $\mib{D}_{\nu}(\xi, \xi'; \mib{R} = 0)$ 
in Fourier transform because this 
belongs to the on-site interaction.
After taking Fourier transform, we diagonalize $\tilde{\mib{D}}_{\nu}(\xi, \xi'; \mib{q})$ for each $\mib{q}$.
We have compared numerical results by varying the number $N$ of neighbors
as $N =32^3$, $64^3$ and $128^3$, and found a reasonable convergence.
The remaining difference is consistent with the asymptotic range dependence of the RKKY interaction given by $R^{-3}$.
We have also checked numerically that $\tilde{\mib{D}}_{\nu}(\xi, \xi'; \mib{q})$ is Hermitian within relative error of order $10^{-5}$.
In the following analysis we use the results with $N = 64^3$.

\subsection{Quadrupolar Interaction}

\begin{figure}[htbp]
\begin{center}
\includegraphics[width=7.5cm, clip]{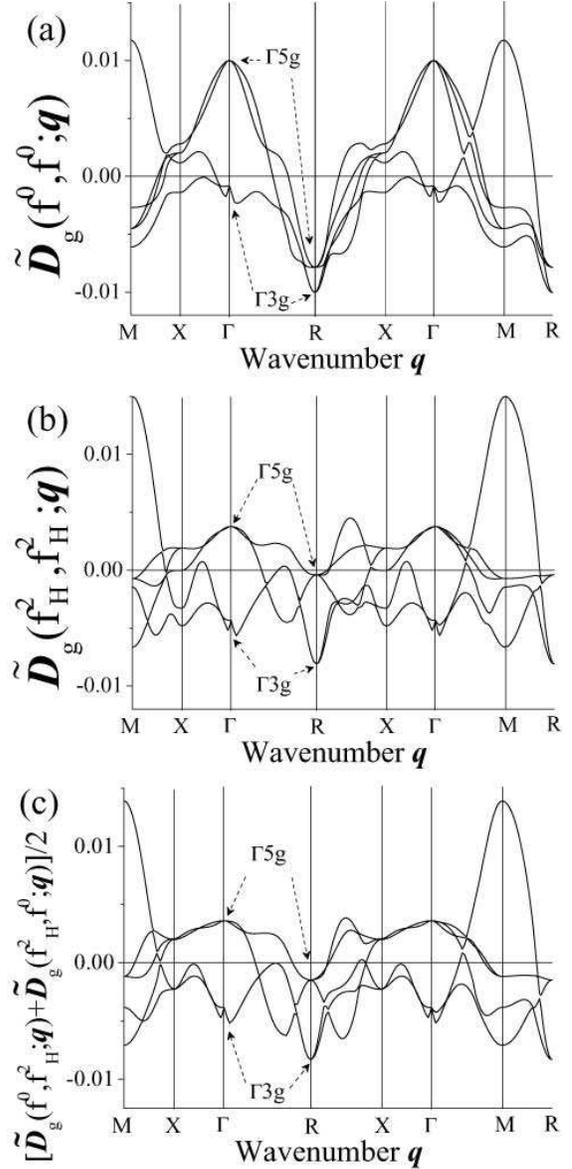}
\caption{ Wavenumber dependence of five eigenvalues of quadrupolar interaction
$\tilde{\mib{D}}_{g}(\xi, \xi'; \mib{q})$.
Results are shown for cases
(a) $f^0$ intermediate states on both sites, 
(b) $f_{\textrm{H}}^2$ intermediate states on both sites,
and (c) $f^0$ on one site and $f_{\textrm{H}}^2$ on the other.
}
\label{FigureQuad}
\end{center}
\end{figure}

We begin with the quadrupolar interactions that 
are even under time reversal in the $\Gamma_8$ manifold. 
Figure \ref{FigureQuad} shows five eigenvalues of quadrupolar couplings $\tilde{\mib{D}}_{g}(\xi, \xi'; \mib{q})$ in the SCBZ with $k_{\textrm{F}}a_q = (3\pi^2)^{1/3}$, which corresponds to one conduction electron per cubic unit cell.
The absolute minimum gives the most stable propagation vector and symmetry of the electronic order.
At high symmetry points such as $\Gamma$ and R, eigenvalues are characterized by irreducible representations of multipoles such as $\Gamma_{3g}$ and $\Gamma_{5g}$.
This is consistent with the symmetry analysis of ref.\citen{Sakai-LCAO}.
For all cases (a), (b) and (c), the minimum occurs at the R point and the most favorable symmetry is $\Gamma_{3g}$.
In the case (a), the eigenvalues of $\Gamma_{5g}$ are comparable
at the R point. 
On the other hand,  $\Gamma_{3g}$ dominates the stability in the cases (b) and (c).
In the previous analysis with only the nearest-neighbor sites,
the $\Gamma_{5g}$ order was found more stable than the $\Gamma_{3g}$ for the case (a)\cite{GS-Kuramoto}.
The difference from the present result  means that interactions between third nearest neighbors or between further distant sites pile up to change the most stable order.  
It remains to identify the reason why the actual CeB$_6$ prefers the $\Gamma_{5g}$ order.  Certainly the actual Fermi surface with three identical pieces should influence the selection.

We note the sharp extrema near the $\Gamma$ point in Fig. \ref{FigureQuad},
which arise due to the Kohn anomaly.
As we have $2k_{\textrm{F}}/(2\pi/a_q) \simeq 0.98$, the location of the Kohn anomaly 
is near the $\Gamma$ point in the reduced zone scheme.
It is well known that the Kohn function (or the static limit of the Lindhard function) does not show a peak at $2k_{\textrm{F}}$, and then the Kohn anomaly appears only as its divergent derivative.
With orbital angular momenta as in eq.(\ref{Hhyb}), however, 
some generalized polarization functions show a hump or a dip around $2k_{\textrm{F}}$ 
as discussed later.

\subsection{Magnetic Interaction}
\begin{figure}[htbp]
\begin{center}
\includegraphics[width=7.5cm, clip]{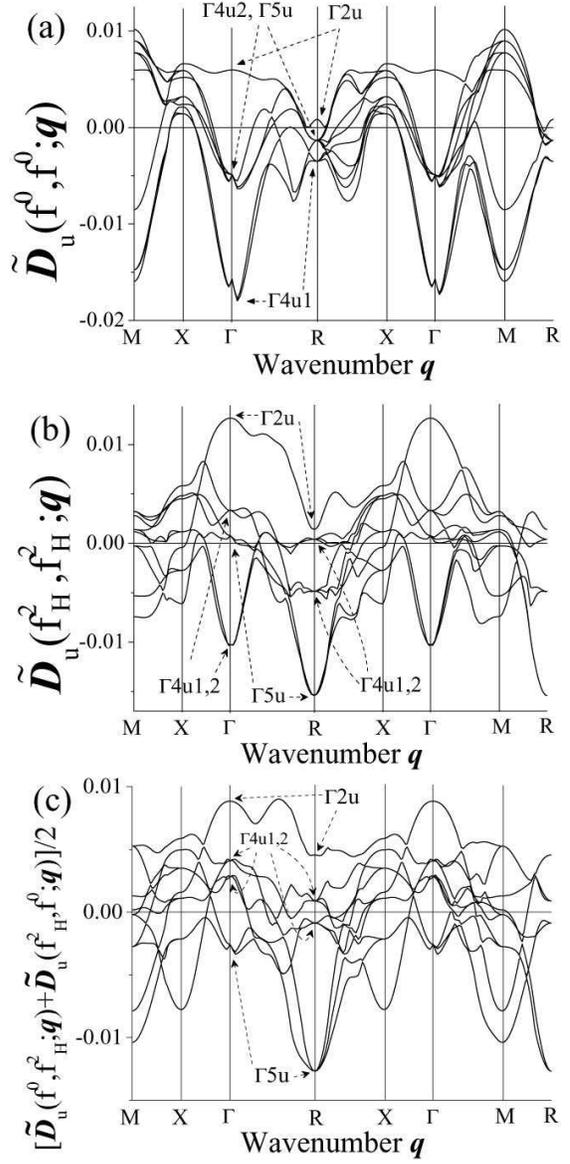}
\caption{Wavenumber dependence of ten eigenvalues of
magnetic multipolar interaction $\tilde{\mib{D}}_{u}(\xi, \xi'; \mib{q})$.
Results are shown for cases
(a) $f^0$ intermediate states on both sites, 
(b) $f_{\textrm{H}}^2$ intermediate states on both sites, and
(c) $f^0$ on one site and $f_{\textrm{H}}^2$ on the other.}
\label{FigureMag}
\end{center}
\end{figure}

Figure \ref{FigureMag} shows eigenvalues of magnetic multipolar couplings $\tilde{\mib{D}}_{u}(\xi, \xi'; \mib{q})$ in the SCBZ with $k_{\textrm{F}}a_q = (3\pi^2)^{1/3}$.  
The minimum gives the propagation vector and symmetry of the favorable magnetic order.
Ten magnetic multipole operators in the $\Gamma_8$ quartet are characterized by irreducible representations $\Gamma_{2u}$, $\Gamma_{4u1}$, $\Gamma_{4u2}$ and $\Gamma_{5u}$.
While the multipolar operators of $\Gamma_{4u1}$ or $\Gamma_{4u2}$ symmetry both have magnetic dipole moment, 
$\Gamma_{2u}$ and $\Gamma_{5u}$ do not contain the dipole character.
The eigenvalues of $\Gamma_{\text{4u2}}$ and $\Gamma_{\text{5u}}$ are degenerate at $\Gamma$ and R points for the case (a).
For the cases (b) and (c), 
the magnetic interactions are mixtures corresponding  to
$\Gamma_{\text{4u1}}$ and $\Gamma_{\text{4u2}}$ symmetries.

For the case (a), the minimum appears near the $\Gamma$ point, and its symmetry is close to $\Gamma_{\text{4u1}}$.
The wavenumber is $\mib{q}_{\text{K}}=(\delta^{''},\delta^{''},\delta^{''})$ with $\delta^{''} = 0.03$ in units of $2\pi/a_q$, which is close to
$1-2k_{\textrm{F}}/(2\pi/a_q) \simeq 0.02$.
Comparing the minima in Figs. \ref{FigureQuad} and \ref{FigureMag}, we find that the corresponding magnetic order is more stable than the quadrupole order.
This means that the incommensurate magnetic order is favored with the present set of parameters.  We emphasize that the Kohn anomaly is responsible for the stability.

\begin{figure}[htbp]
\begin{center}
\includegraphics[width=8.0cm, clip]{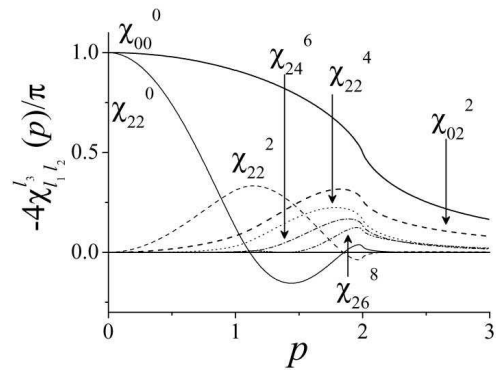}
\end{center}
\caption{Radial parts of generalized polarization functions $\chi_{l_1 l_2}^{l_3}(p)$ with $p=|\mib{q}|/k_{\textrm{F}}$.}
\label{FigureChi}
\end{figure}

With orbital angular momenta as in eq.(\ref{Hhyb}), generalized polarization functions are given by
\begin{align}
&\sum_{\mib{k} \mib{k}'}\delta_{\mib{k}-\mib{k}',\mib{q}}
\frac{f(\epsilon_{k})-f(\epsilon_{k'})}{\epsilon_{k}-\epsilon_{k'}}
Y_{l_{1} m_{1}}(\hat{\mib{k}})Y_{l_{2} m_{2}}(\hat{\mib{k}}')
\nonumber \\
&= \frac{1}{\epsilon_{\textrm{F}}}\left( \frac{k_{\textrm{F}} a_q}{\pi} \right)^3
\sum_{l_3 m_3}i^{-l_{1}+l_{2}+l_3}Y_{l_3 m_3}(\hat{\mbox{\boldmath$q$}})
\chi_{l_{1}l_{2}}^{l_3}(|\mbox{\boldmath$q$}|/k_{\textrm{F}})
\nonumber \\ & \qquad \times
\int \mathrm{d}\Omega \,
Y_{l_3 m_3}^\ast(\Omega)Y_{l_{1}m_{1}}(\Omega)Y_{l_{2} m_{2}}(\Omega), \label{ffYY}
\end{align}
where $f(\epsilon)$ is the Fermi distribution function, $\epsilon_{k}$ is the energy dispersion of free electrons, and $l_{1},l_{2}$ are even integers with $0 \le l_{1}, l_{2} \le 6$.
Note that 
terms with $l_{3} \ne 0$ give anisotropic functions in the wavenumber space.
In addition, angular momentum
in eq.(\ref{ffYY}) makes the $|\mib{q}|$-dependence of the polarization function rather different from the ordinary Kohn function.
For understanding the characteristic behaviour for finite $l_1$ and $l_2$,
we consider the hypothetical RKKY interaction 
angular momentum $l \gg 1$ and with $m = 0$.
The corresponding polarization function is given by
\begin{align}
&\sum_{\mib{k} \mib{k}'}\delta_{\mib{k}-\mib{k}',\mib{q}}
\frac{f(\epsilon_{k})-f(\epsilon_{k'})}{\epsilon_{k}-\epsilon_{k'}}
|Y_{l 0}(\hat{\mib{k}})|^2
|Y_{l 0}(\hat{\mib{k}}')|^2 , \label{hypoChi}
\end{align}
where $Y_{l 0}(\hat{\mib{k}})$ corresponds to $Y_{3m}(\hat{\mib{k}})$ in eq.(\ref{Hhyb}).
The significant contribution in summation over $\hat{\mib{k}}$ and $\hat{\mib{k}}'$ 
comes only from  directions around the $z$-axis.
This is because the spherical harmonics 
with $l \gg 1$ behaves asymptotically as \cite{Varshalovich}
\begin{equation}
Y_{l 0}(\theta, \phi) \approx 
\begin{cases}
\sqrt{\frac{2l+1}{4\pi}}[1-\frac{l(l+1)}{2} \theta^2] \qquad {\rm for }\quad 0 \leq \theta \leq \varepsilon (\ll 1), \\
\dfrac{\cos[(l+\tfrac{1}{2})\theta - \tfrac{\pi}{4}]}{ \pi \sqrt{\sin{\theta}} }
 \qquad {\rm for} \quad l^{-1} \ll \theta \ll \pi - l^{-1}, \\
(-1)^{l}\sqrt{\frac{2l+1}{4\pi}}[1-\frac{l(l+1)}{2} (\pi \!- \! \theta)^2] \quad {\rm for} \, \pi-\varepsilon \leq  \theta \leq \pi, \\
\end{cases}
\end{equation}
and is sharply peaked at $\theta =0,\pi$.
Then the polarization function acquires a one-dimensional character 
with $l \gg 1$ and is strongly peaked at $\mib{q}=(0,0,\pm 2k_{\textrm{F}})$.
This anisotropic behaviour should be reflected in the terms with $l_{3} \ne 0$ in eq.(\ref{ffYY}) around $|\mib{q}| = 2k_{\textrm{F}}$.
Figure \ref{FigureChi} shows radial parts of generalized polarization functions $\chi_{l_1 l_2}^{l_3}(p)$ with $p = |\mib{q}|/k_{\textrm{F}}$.
Except for $\chi_{00}^{0}(p)$, which is the Kohn function, 
these functions have a hump or a dip at $|\mib{q}| \simeq 2k_{\textrm{F}}$. 
They can accumulate to give a sharp extremum near $|\mbox{\boldmath$q$}| = 2k_{\textrm{F}}$ in the intersite interaction.

The wavenumber for the Kohn anomaly in the SCBZ
is 
given by 
$\delta_{\textrm{K}}\equiv \!1\!-\!2k_{\textrm{F}}/(2\pi/a_q)$ in units of the reciprocal lattice.
To be precise, $\delta_{\textrm{K}}$ differs to some extent from $\delta^{''}$ obtained numerically, 
because each extremum in $\chi_{l_1 l_2}^{l_3}(p)$ is off from $p = 2$ in general.
We note that 
the magnetic interaction takes another minimum at the M points (1/2, 1/2, 0), (0, 1/2, 1/2) and (1/2, 0, 1/2), and its symmetry is close to $\Gamma_{4u2}$.
The anisotropy of the exchange is small in the case (a), but is large in cases (b) and (c).
This is consistent with the results obtained in the previous paper \cite{GS-Kuramoto}.

For cases (b) and (c), the minimum occurs at the R point, and its symmetry is $\Gamma_{\text{5u}}$.
This means that the antiferro octupolar ordering is stabilized.
The wavenumber $\mib{q}_{\text{K}}$ also gives a local minimum, but the anomaly is weaker than the case (a).
In the case (c), 
the dipole interaction is insignificant for all wavenumbers.
This is consistent with the smallness of dipolar couplings in the nearest and next nearest neighbors as derived in the previous paper \cite{GS-Kuramoto}.

\subsection{Other Choices of Lowest CEF States}
So far multipolar interactions are calculated with $\Gamma_8$ quartet as the lowest CEF states.
In this subsection, we 
choose other states as the lowest CEF states: (i) $j_1 = 5/2, m = \pm 1/2$ doublet and (ii)$\Gamma_7$ doublet of $j_1 = 5/2$.
Here only the 4$f^0$ is considered as an intermediate state.

\begin{figure}[htbp]
\begin{center}
\includegraphics[width=7.5cm, clip]{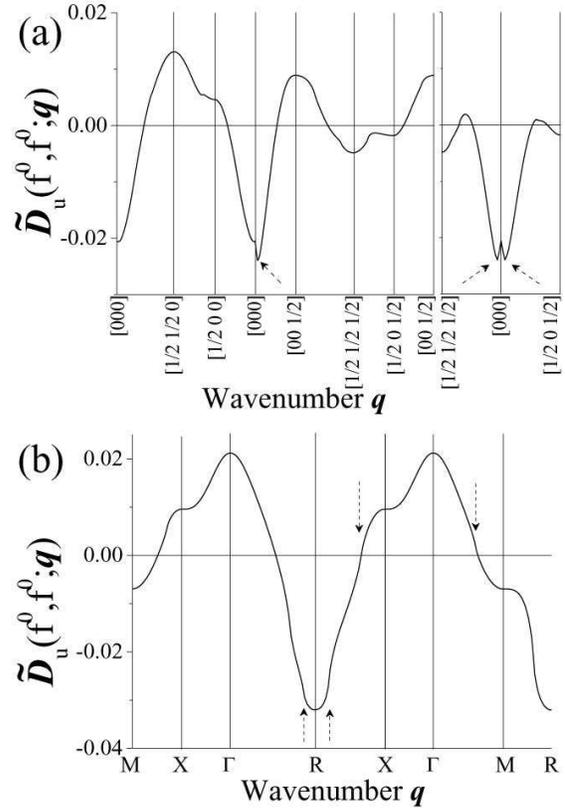}
\caption{Wavenumber dependence of magnetic interaction $\tilde{\mib{D}}_{u}(f^0, f^0; \mib{q})$ with $k_{\textrm{F}}a_q = 0.98\pi$.
Results are shown for the lowest CEF states 
(a) $j_1 = 5/2, m = \pm 1/2$ doublet, and 
(b) $\Gamma_7$ doublet.
Arrows indicate the location of the Kohn anomaly.
}
\label{FigureCEF}
\end{center}
\end{figure}

We first take the dimensionless wavenumber $k_{\textrm{F}}a_q = 0.98\pi$, which corresponds to one conduction electron per cubic unit cell.
Figure \ref{FigureCEF}(a) shows magnetic dipolar interaction for the $m = \pm 1/2$ doublet. 
The wavefunction has uniaxial anisotropy along [001] direction.  Hence the results are shown for a BZ accounting for 
the corresponding anisotropy. 
With 4$f^0$ intermediate states, the dipolar interaction between 
the Kramers doublets is a scalar. 
Around the $\Gamma$ point, the Kohn anomaly appears and gives the minimum.
The preferred wavenumber is the same as
that for the $\Gamma_8$ quartet as the lowest CEF states.
We note that the Kohn anomaly is conspicuous along [001], while almost invisible along [100].  Such strong anisotropy comes from the highly elongated shape of the wave function along [001].  

\begin{figure}[htbp]
\begin{center}
\includegraphics[width=7.5cm, clip]{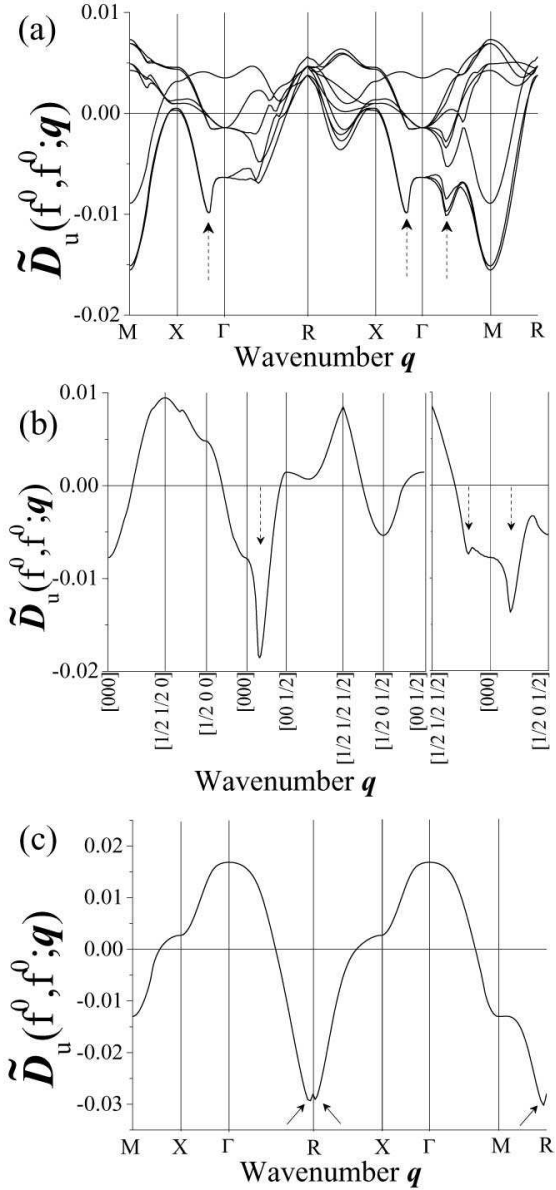}
\caption{ Wavenumber dependence of magnetic interaction $\tilde{\mib{D}}_{u}(f^0, f^0; \mib{q})$ with $k_{\textrm{F}}a_q = 0.86\pi$.
Results are shown for the lowest CEF states,
(c) $\Gamma_8$ quartet,
(b) $j_1 = 5/2, m = \pm 1/2$ doublet, and
(c) $\Gamma_7$ doublet.
Arrows indicate the location of the Kohn anomaly.
}
\label{FigureCEF086}
\end{center}
\end{figure}

Figure \ref{FigureCEF}(b) shows magnetic dipolar interaction for the $\Gamma_7$ doublet where the Kohn anomaly is not obvious.
However, the derivative of $\tilde{\mib{D}}_{u}(f^0, f^0; \mib{q})$ with respect to $\mib{q}$ seems to diverge at $(\delta_7^{(1)}, \delta_7^{(1)}, \delta_7^{(1)})$ with $\delta_7^{(1)} \simeq 0.43$.
This is interpreted as the three-dimensional Kohn anomaly.  Namely, 
taking into account the reciprocal lattice vector along [111], 
we obtain the Kohn anomaly at
$\delta_{\textrm{K}}' = 1 - k_{\textrm{F}}a_q/\sqrt{3}\pi$.
Here the dimensionless wavenumber $k_{\textrm{F}}a_q$ is 0.57$\sqrt{3}\pi$ and gives $\delta_{\textrm{K}}' \sim 0.43$.
Three more anomalies can be found at $(\delta_7^{(2)}, \delta_7^{(2)}, 0)$ with $\delta_7^{(2)} \sim 0.32, 
(\delta_7^{(3)}, \delta_7^{(3)}, 1/2)$ with $\delta_7^{(3)} \sim 0.20$, 
and $(\delta_7^{(4)}, \delta_7^{(4)}, 1/2)$ with $\delta_7^{(4)} \sim 0.39$.
These wave numbers are interpreted as intercepts 
of spheres with its radius $2k_{\textrm{F}}$ 
at high symmetry axes in the irreducible BZ.
The spheres are centered at reciprocal lattice points [110], [100] and [111], respectively,

In order to study the effect of band structures in the simplest way, we now change the Fermi wavenumber.
Namely we take $k_{\textrm{F}}a_q = 0.86\pi \simeq 0.497\sqrt{3}\pi$, and show the results in Fig.\ref{FigureCEF086}.
For comparison with the main case of $k_{\textrm{F}}a_q = 0.98\pi$, 
we also show in Fig.\ref{FigureCEF086}(a) the magnetic multipolar interaction with the $\Gamma_8$ quartet. 
Figure \ref{FigureCEF086}(b) shows the magnetic dipolar interaction for the $m=\pm 1/2$ doublet.
The Kohn anomaly is found at $(0, 0, \delta^{'''})$ with $\delta^{'''} = 0.17$,
which is close to $1-k_{\textrm{F}}a_q/\pi = 0.14$.
In the case of $\Gamma_7$ doublet shown in Fig.\ref{FigureCEF086}(c), on the other hand, 
Kohn anomalies in the magnetic dipolar interaction
are found around the R point.
The minimum is at 
$(\delta_7^{(5)}, \delta_7^{(5)}, \delta_7^{(5)})$ 
with $\delta_7^{(5)} \simeq 0.48$.
Since $2k_{\textrm{F}}\simeq 0.994 \sqrt{3}\pi/a_q$ is almost the same as the distance between $\Gamma$ and R points,
the Kohn anomaly can arise near the R point.
From the large difference between the results for each doublet,
we see that the anisotropy of the lowest CEF states sensitively
influences the wavenumber dependence of the interaction, and the character of the Kohn anomaly.

\section{Discussion}
\subsection{Comparison with real-space results}
We now compare 
between the present result and the previous one \cite{GS-Kuramoto} which takes into account only the nearest and the next-nearest neighbor interactions.
The previous study in addition neglects possible mixing of different point-group representations in the intersite interaction.
As clarified by Sakai {\it et al}.\cite{Sakai-LCAO}, the mixing becomes minimum at high symmetry points such as R in the SCBZ.  
Thus the result can be compared with previous study if we look at only $\Gamma$ and R. 
However $\Gamma_{4u1}$ and $\Gamma_{4u2}$ representations still mix at $\Gamma$ and R points as seen in Fig \ref{FigureMag}(b) and (c).
Concerning the time-reversal odd interactions,
occurrence of ferromagnetic dipolar interaction and antiferromagnetic $\Gamma_{5u}$ octupolar interaction is consistent with the previous result.  
In the present case, however, it is difficult to judge whether the interactions have a significant anisotropic, or ``pseudo-dipolar'' character.   
If we look at only the R point in Fig.\ref{FigureMag}(b) and (c), 
eigenvalues of $\Gamma_{5u}$ appear as triply degenerate.
The same degeneracy also applies to the eigenvalues of $\Gamma_{4u}$ representations.
The pseudo-dipole character can only be seen by gathering information at various wavenumbers.

On the other hand, the implication for quadrupole order is different between the previous result \cite{GS-Kuramoto} and the present one.
Namely we found in the previous study that the nearest-neighbor interaction prefers staggered quadrupole order with $\Gamma_{5g}$ symmetry to $\Gamma_{3g}$, because the interaction between $\Gamma_{5g}$ moments is antiferro-type but that between $\Gamma_{3g}$ moments includes both ferro- and antiferro-type couplings.
In the present study, it is found that many distant pairs interact and make $\Gamma_{3g}$ more favorable, although the R point remains the most favorable wavenumber. 
The relative stability of $\Gamma_{3g}$ against $\Gamma_{5g}$ is fragile if $f^0$ is the dominant intermediate state.  As the $f_{\text{H}}^2$ intermediate states become important, $\Gamma_{3g}$ becomes more stable.
It is the effect of many distant neighbors that makes the difference from previous results.   Thus there is no inconsistency between both calculations.

\subsection{Relevance to real Ce materials such as CeB$_6$ and CeB$_2$C$_2$}
In CeB$_6$,  it is known that the quadrupole order sets in with the $\Gamma_{5g}$ symmetry.  It is not known, however, to what extent it is stable against another candidate with $\Gamma_{3g}$.  Obviously the present model is oversimplified to be applied to CeB$_6$, especially because of the plane-wave conduction band with a single Fermi surface.
We are now using more realistic band structures to derive 
intersite interactions. 

It is remarkable that the plane-wave model prefers $\Gamma_{5u}$ octupole order if the $f_{\text{H}}^2$ intermediate states are dominant.   
The result is encouraging to understand the microscopic origin of possible octupole order in phase IV of Ce$_{x}$La$_{1-x}$B$_6$ with $x\sim 0.75$ \cite{Kubo-Kuramoto,Tayama,Morie}.
It should be noted that the $f^0$ intermediate state does not favor $\Gamma_{5u}$ at all.
Therefore use of realistic energy band is again important to understand the microscopic mechanism.

We now turn to CeB$_2$C$_2$ where the incommensurate magnetic order is observed.  Since the Fermi surface is essentially an ellipsoid, our model may apply more directly to this material.
The lowest CEF states of CeB$_2$C$_2$ is considered to be the $m = \pm 1/2$ doublet, which is split off from $\Gamma_8$ by the tetragonal CEF.
We have seen that the strong Kohn anomaly appears near the $\Gamma$ point of the Brillouin zone.
In view of the dHvA experiment of LaB$_2$C$_2$ \cite{Watanuki1},
let us take $2k_{\textrm{F}}= 0.86(2\pi/a_q)$ which gives $\delta_{\textrm{K}} =0.14$. 
The Kohn anomalies in our calculation arise at $(\delta^{'''}, 0, 0)$ and $(\delta^{'''}, \delta^{'''}, 0)$ with $\delta^{'''} \simeq 0.17$ in the SCBZ.
Mapping to the QCBZ, the relevant 
wavenumbers are $(\delta^{'''}, 0, 0)$, $(\delta^{'''}, \delta^{'''}, 0)$, $(\delta^{'''}, \delta^{'''}, \delta^{'''})$ and their cyclic rotations.  
These values are in good correspondence with experimental value of $\delta = 0.16$ already cited in \S 1.
Actually the wavenumbers giving large intersite interaction form a ridge in the BZ, and the points quoted above are part of the ridge.
We note that characteristic wavenumbers of RB$_2$C$_2$ compound, $(1 \pm \delta, \pm \delta, 0)$ in the QCBZ \cite{Ohoyama4}, are close to intercepts of the sphere with radius $2k_{\textrm{F}} = 0.86(2\pi/a_q)$ at the BZ boundary.
Thus we interpret  incommensurate order found in RB$_2$C$_2$ 
in terms of the Kohn anomaly.

In concluding this section 
we mention a theoretical study on 
the RKKY interaction between Pr nuclei\cite{Nagasima}, 
where Kohn anomalies with a peak structure is found, although the authors do not remark on this feature.

\section{Summary}
We have calculated the RKKY interaction in the Anderson lattice with orbital degeneracy in the simple cubic lattice, and clarified
the relation between the multipolar ordering structure and the Fermi surface.
For intermediate states, we have considered both 4$f^0$ and 4$f^2$ Hund's-rule ground states 
as in the previous paper.
Our simplified model gives an incommensurate magnetic structure as the most stable structure, 
for which orbital angular momentum of 4$f$ electrons plays a crucial role.
The propagation vector for incommensurate magnetic structure is determined by the Kohn anomaly 
at $2k_{\textrm{F}}$ which can be shifted by reciprocal lattice vectors.
With one conduction electron per cell, this anomaly can arise near the $\Gamma$ point.
If we take the Fermi wavenumber from the dHvA experiment of LaB$_2$C$_2$,
the propagation vector is in good correspondence with that of incommensurate magnetic structure in CeB$_2$C$_2$. 
This result should be relevant in understanding the 
incommensurate magnetic structures observed in RB$_2$C$_2$ in general, 
but effects of anisotropy should be taken into account for more quantitative study.

The anisotropy of CEF states 
influences the wavenumber dependence of interaction, i.e., the most favorable structure and the wavenumber of Kohn anomaly.
The propagation vector is unchanged 
whether the lowest CEF state is $\Gamma_8$ quartet or $m = \pm 1/2$ doublet for both intermediate states 4$f^0$ and 4$f_{\text{H}}^2$. 
The most stable ordering structure in the case of 4$f^0$ intermediate state is 
the incommensurate magnetism, while the 4$f_{\text{H}}^2$ intermediate states give a staggered octupolar order.
It is still an interesting open problem to clarify the relation between the multiplet structure of $f$ shell other than Ce and the order parameters. 
\section*{Acknowledgement}
One of us (G. S.) would like to thank Dr. T. Matsumura for useful discussions about magnetic structures of RB$_2$C$_2$.
\appendix
\section{Generalized Polarization Functions}
In this section, we 
present analytic results of generalized polarization functions.
The radial parts of generalized polarization functions in the main text are given by
\begin{align}
\chi_{l_{1}l_{2}}^{l_3}(p)
&\equiv
\int_0^{\infty} \frac{\mathrm{d}z}{z^2}j_{l_3}\left(pz\right)
\int_0^z t^2 \mathrm{d}t \int_0^{\infty} \frac{{t'}^2 \mathrm{d}t'}{t^2 - {t'}^2}j_{l_{1}}(t)j_{l_{2}}(t')
\nonumber \\ 
&\equiv \int_0^{\infty} \frac{\mathrm{d}z}{z^2}j_{l_3}\left(pz\right)q(l_{1},l_{2};z).
\label{chilll}
\end{align}
where $j_l(x)$ is the spherical Bessel function of order $l$, and $q(l_{1},l_{2};z)$'s are tabulated in ref.\citen{GS-Kuramoto}.
Hereafter we consider even $l_3$, since otherwise the integration about the solid angle $\Omega$ vanishes in eq.(\ref{ffYY}).

From the analysis around $z = 0$, we find that $z^{-2}q(l_{1},l_{2};z) $ behaves as $z^{\textrm{min}(l_1,l_2)+1}$, and that $j_{l_3}(z)$ includes terms like $z^{-l_3 - 1}\sin(z)$ and $z^{-l_3}\cos(z)$.
Therefore in reducing eq.(\ref{chilll}) to Fourier sine and cosine transforms, the following three cases are considered: (i)$l_3 \le \min(l_{1},l_{2})$, (ii)$l_3 = l_{1} + l_{2}$, (iii)otherwise.
In the case (i), the calculation is straightforward once $z^{-2}q(l_{1},l_{2};z) $ is divided into the sum of regular forms at $z = 0$, {\it e.g.}
\begin{align}
\frac{2}{\pi z^2} q(0, 0;z)
=& \frac{1}{2z}[\cos(2z) - 1] - \frac{1}{4z^2}[\sin(2z) - 2z].
\end{align}
In the case (ii), we transform the right hand side of eq.(\ref{chilll}) as follows:
\begin{equation}
p \int_0^{\infty}\frac{\mathrm{d}z}{z^2}j_{l_3}(z)q(l_{1},l_{2};z/p), 
\end{equation}
and consider the transform of $j_{l_3}(z)/z^2$. 
This calculation is almost the same as that for the case (i), 
because $q(l_{1},l_{2};z)/z^2$ includes terms like $z^{-l+1}\cos(z), z^{-l}\sin(z)$ with $0 \le l \le (l_1+l_2)$,
and $\text{Si}(z)/z^2$ with the sine integral Si$(z)$.
In the case (iii), we combine the above two cases
to obtain $\chi_{l_{1}l_{2}}^{l_3}(p)$.
Relevant 
transforms are as follows:
\begin{align}
&\int_0^\infty \mathrm{d}z\, \frac{\sin(pz)}{z^{2m+3}}
\left[
\sin(z)-\sum_{n=0}^{m} (-1)^n \frac{z^{2n+1}}{(2n+1)!}
\right] \nonumber \\
&=\frac{(-1)^{m+1}}{(2m+2)!}
\left\{
\frac{1}{2}
	\left[(p+1)^{2m+2} \log |1+\frac{1}{p}|
	\right. \right. \nonumber \\ & \qquad \qquad \qquad \quad \left. \left.
	-(p-1)^{2m+2} \log |1-\frac{1}{p}|
	\right]
\right. \nonumber \\
& \quad \left.
-\sum_{k=0}^{m+1}
\begin{pmatrix}
2m+2 \\
2k
\end{pmatrix}
\sum_{l \ge 0}^{k-1} \frac{p^{2k-2l-1}}{2l+1}
	\right. \nonumber \\ &  \quad \left. 
+\sum_{k=0}^m
\begin{pmatrix}
2m+2 \\
2k+1
\end{pmatrix}
\sum_{l \ge 0}^{k-1} \frac{p^{2k-2l-1}}{2l+2} \right\}, \label{SinNormalSinTransform}
\end{align}
\begin{align}
&\int_0^\infty \mathrm{d}z\, \frac{\sin(pz)}{z^{2m+2}}
\left[
\cos(z)-\sum_{n=0}^{m} (-1)^n \frac{z^{2n}}{(2n)!}
\right] \nonumber \\
&=\frac{(-1)^{m+1}}{(2m+1)!}
\left\{
\frac{1}{2}
\left[
(p+1)^{2m+1}\log|1+\frac{1}{p}|
	\right. \right. \nonumber \\ & \qquad \qquad \qquad \quad \left. \left.
+(p-1)^{2m+1}\log|1-\frac{1}{p}|
\right]
\right. \nonumber \\
& \quad \left.
-\sum_{k=0}^{m}
\begin{pmatrix}
2m+1 \\
2k
\end{pmatrix}
\sum_{l \ge 0}^{k-1}
\frac{p^{2k-2l-1}}{2l+1}
	\right. \nonumber \\ &  \quad \left. 
+\sum_{k=0}^{m}
\begin{pmatrix}
2m+1 \\
2k+1
\end{pmatrix}
\sum_{l \ge 0}^{k-1} \frac{p^{2k-2l-1}}{2l+2}
\right\}, \label{CosNormalSinTransform}
\end{align}
\begin{align}
&\int_0^\infty \mathrm{d}z \, \sin(pz) \left(\frac{\mathrm{d}}{\mathrm{d}z} \right)^{2n} \frac{ \textrm{Si}(z)}{z}
 \nonumber \\
&= -p \int_0^\infty \mathrm{d}z \, \cos(pz) \left(\frac{\mathrm{d}}{\mathrm{d}z} \right)^{2n-1} \frac{ \textrm{Si}(z)}{z}, \label{RecurrenceSin} \\
&\int_0^\infty \mathrm{d}z \, \cos(pz) \left(\frac{\mathrm{d}}{\mathrm{d}z}\right)^{2n+1} \frac{ \textrm{Si}(z)}{z}
 \nonumber \\
&=\frac{(-1)^{n+1}}{(2n+1)^2}
+p \int_0^\infty \mathrm{d}z \, \sin(pz) \left(\frac{\mathrm{d}}{\mathrm{d}z} \right)^{2n} \frac{ \textrm{Si}(z)}{z}, \label{RecurrenceCos}
\end{align}
where $(_r^n)$ is 
the binomial coefficient. 

In this way analytic result of $\chi_{l_1l_2}^{l_3}(p)$ is obtained, and some of the results are tabulated in Table~\ref{chitable}.
In Table~\ref{chitable} we define functions $f(x)$ and $g(x)$ by
\begin{align}
f(x) &\equiv (x^2-1)\log\left| \frac{1+x}{1-x} \right|, \\
g(x) &\equiv \int_0^{\infty} \frac{\mathrm{d}z}{z} \sin(xz) \text{Si}(z) \nonumber \\
&= 
\begin{cases}
\frac{\pi^2}{4} - \frac{1}{2}\left[ L_2(x) - L_2(-x)\right] & 0 < |x| < 1 \\
\frac{1}{2}\left[ L_2(1/x)-L_2(-1/x) \right] & |x| > 1
\end{cases},
\end{align}
where $L_2(x)$ is the Euler's dilogarithm,\cite{Gradshteyn}
\begin{equation}
L_2(x) = - \int_0^x \mathrm{d}z \frac{\log(1-z)}{z}.
\end{equation}
We note that $\chi_{00}^{0}(p)$ is the 
Kohn function,
and that $\chi_{ll}^{0}(p=0)$ converges to the constant value, $-\pi/4$.
\begin{align}
&
\chi_{ll}^{0}(p = 0)
\nonumber \\ &
= \lim_{T \to 0} \frac{1}{k_{\textrm{F}}}\int_0^{\infty} \!k^2 \mathrm{d}k
\int_0^{\infty} {k'}^2 \mathrm{d}k'
\nonumber \\
& \qquad \qquad \times
 \int_0^{\infty} \!x^2 \mathrm{d}x \frac{f(\epsilon_k)-f(\epsilon_{k'})}{k^2 - {k'}^2}
j_l(kx)j_l(k'x) \nonumber \\
&= \lim_{T \to 0} \frac{\pi}{4 k_{\textrm{F}}} \int_0^{\infty} k \frac{\mathrm{d} f(\epsilon_k)}{\mathrm{d} k} \mathrm{d}k \qquad = -\frac{\pi}{4}.
\end{align}
Here, we have used the following relation,\cite{Arfken}
\begin{equation}
\frac{2a^2}{\pi} \int_0^\infty j_l(ax)j_l(bx)x^2 \mathrm{d}x = \delta(a-b).
\end{equation}
This is related to the asymptotic behaviour of range functions in real space, \cite{Coqblin}
\begin{equation}
z^{-4}q(l_{1},l_{2}; z) \simeq (-1)^{\tfrac{l_{1}+l_{2}}{2}}\frac{z \cos z - \sin z}{z^4} \qquad (z \to \infty).
\end{equation}
In general $\chi_{l_{1}l_{2}}^{l_3}(p)$ is an oscillatory function for $0 < p < 2$, and are slowly varying for $p > 2$.
We have checked the results in Fig.\ref{FigureChi} or in Table~\ref{chitable} 
numerically by spherical Bessel transforms \cite{Farazdel}.
\begin{table}
\caption{
Polarization functions with orbital degeneracy $2\chi_{l_1 l_2}^{l_3}(p)/\pi$. The 
definitions of $f(x)$ and $g(x)$ are given in the main text.
This table gives results for max$(l_1, l_2) \le 2$.}
\label{chitable}
\begin{tabular}{cccrrrr} \hline
$l_1$ & $l_2$ & $l_3$ & $p^{-5}f(p)$ & $p^{-3}f(p)$ & $p^{-1}f(p)$ & $pf(p)$ \\ \hline
0 & 2 & 2 & \, & $ 3/64 $ & $- 9/64 $ & \, \\ \hline
2 & 2 & 0 & \, & \, & $ 3/16 $ & $ 3/16 $ \\
2 & 2 & 2 & \, & $- 3/64 $ & $ 3/64 $ & $- 3/32 $ \\
2 & 2 & 4 & $ 35/768 $ & $- 55/768 $ & $- 1/768 $ & $ 3/256 $ \\ \hline
\end{tabular}

\begin{tabular}{cccrr} \hline
$l_1$ & $l_2$ & $l_3$ & $p^{-1}f(p/2)$ & $pf(p/2)$ \\ \hline
0 & 2 & 2 & $1/8$ & \, \\ \hline
2 & 2 & 0 & $- 1/4 $ & $- 3/8 $ \\
2 & 2 & 2 & $- 1/8 $ & $ 3/16 $ \\
2 & 2 & 4 & $- 3/32 $ & $- 3/128 $ \\ \hline
\end{tabular}

\begin{tabular}{cccrrrr} \hline
$l_1$ & $l_2$ & $L$ & $p g(p/2)$ & $p g(2/p)$ & $p g(p)$ & $p g(1/p)$ \\ \hline
0 & 2 & 2 & \, & $ 3/8$ & \, & $-3/8 $ \\ \hline
2 & 2 & 0 & $ 3/2$ & \, & $-3/2 $ & \, \\
2 & 2 & 2 & $-3/8$ & \, & $ 3/8 $ & \, \\
2 & 2 & 4 & \, & $1/16$ & \, & $-1/16 $ \\ \hline
\end{tabular}

\begin{tabular}{cccrrr} \hline
$l_1$ & $l_2$ & $l_3$ & $p^{-4}$ & $p^{-2}$ & $p^{0}$ \\ \hline
0 & 2 & 2 & \, & $- 3/32 $ & $ 7/32$ \\ \hline
2 & 2 & 0 & \, & \, & $- 5/8 $ \\
2 & 2 & 2 & \, & $ 3/32 $ & $- 1/32 $ \\
2 & 2 & 4 & $- 35/384 $ & $ 235/1152 $ & $ 5/384$ \\ \hline
\end{tabular}
\end{table}

\end{document}